\documentclass[twocolumn,showpacs,preprintnumbers]{revtex4}
\usepackage{graphicx}
\usepackage{dcolumn}
\usepackage{bm}

\begin{document}

\title{Unconventional superfluidity induced by spin-orbital coupling in a
polarized two-dimensional Fermi gas}
\author{J.-N. Zhang$^{1}$, Y.-H. Chan$^{2,1}$ and L.-M. Duan$^{2,1}$}
\affiliation{$^{1}$Center for Quantum Information, IIIS, Tsinghua University, Beijing,
China}
\affiliation{$^{2}$Department of Physics and MCTP, University of Michigan, Ann Arbor,
Michigan 48109, USA }

\begin{abstract}
We show the spin-orbital coupling induced by an artificial light-induced gauge field can fully restore superfluidity suppressed by population
imbalance in a two-dimensional (2D) Fermi gas, leading to unconventional
superfluid states either with topological Majorana fermion excitations or
showing a novel mixture of triplet pairing with spin-up (down) components
respectively in the $p_{x}\pm ip_{y}$ pairing channels. We self-consistently
calculate the zero temperature phase diagram at the BCS\ side of Feshbach
resonance and show that the phase transitions between different superfluid
states can be revealed through measurement of the in-situ density profile of
the 2D atomic cloud in a weak global trap.
\end{abstract}

\maketitle

Ultracold atomic gas provides an ideal platform to study superfluid states
under tunable configurations. There has been great interest in looking for
unconventional superfluid states beyond the conventional s-wave
superfluidity observed in experiments \cite{1}. Population imbalance
provides a mechanism to suppress the s-wave superfluid state and can lead to
novel superfluid phase, such as the FFLO\ state \cite{2}. Polarized Fermi
gas with tunable population imbalance has been studied extensively both
experimentally and theoretically \cite{1,2}. The FFLO\ state, unfortunately,
is typically fragile and hard to observe experimentally \cite{1,2}.
Increasing population imbalance usually destroys the superfluid state
instead of leading to novel superfluidity.

The spin-orbital (SO) coupling induced by an artificial gauge field emerges
as a powerful control method for ultracold atomic gas \cite{3,4,5}. Various
proposals have been made to realize SO coupling in ultracold atomic gas \cite%
{3,4} and remarkable experimental progress has been reported to demonstrate
the artificial gauge field \cite{5}. Motivated by this progress, strong
interest arises recently in studying the three-dimensional ultracold Fermi
gas under artificial SO coupling \cite{6}. It has been noted that the SO
coupling can enhance superfluidity for this system \cite{6}.

In this paper, we show that interesting and exotic superfluid states arise
in two-dimensional (2D) Fermi gas from interplay of population imbalance and
SO coupling. Without SO coupling, population imbalance above a critical
value suppresses superfluidity, leading to normal phases. We show that SO
coupling fully restores superfluidity in 2D, but the pairing is not in the
conventional $s$-wave channel any more and the resultant superfluid phases
have more exotic properties. Under large polarization, one of the spin
components dominates with the pairing in the $p_{x}+ip_{y}$ channel. Such a
phase supports non-abelian topological excitations by trapping a Majorana
Fermion in its vortex core \cite{7}. This phase has been predicted before
for a 2D Fermi gas \cite{7,8}, and the contribution of our calculation of a
self-consistent phase diagram is to determine the stability region of such a
phase under real physical parameters, which is important for experimental
observation. Furthermore, our calculation predicts a new kind of
unconventional superfluid phase under intermediate polarization, with
spin-dependent pairing in the $p_{x}\pm ip_{y}$ channels. In this phase, the
majority (minority) components pair up respectively in the $p_{x}+ip_{y}$ ($%
p_{x}-ip_{y}$) channels. We characterize the transition order between
different superfluid states, and show through explicit calculation that
these transitions can be revealed by observation of the singularity points
of the density profile of a 2D atomic cloud in a weak global trap, which can
be measured in experiments through the in-situ imaging \cite{9}.

In experiments, one realizes the 2D Fermi gas by applying a strong optical
trapping potential (or an optical lattice) to ultracold atoms along the $z$%
-direction. Near the potential minimum where the atoms are located, the
potential is well approximated by a harmonic trap with $V_{z}\left( z\right)
=m\omega _{z}^{2}z^{2}/2,$where $m$ denotes the atomic mass and $\omega _{z}$
is the trapping frequency. At the BCS side of the Feshbach resonance, the
transverse wave function for the atoms is given by the ground state of $%
V_{z}\left( z\right) $, and the atomic collision can be characterized by an
effective 2D interaction~\cite{10}. With light induced Rashba-type of SO
coupling \cite{3,4,5,6}, the Hamiltonian for the system can be described as
\begin{eqnarray}
H &=&\!\!\!\sum_{{\mathbf{k}},\sigma }\xi _{{\mathbf{k}},\sigma }\hat{a}_{{%
\mathbf{k}},\sigma }^{\dag }\hat{a}_{{\mathbf{k}},\sigma }\!\!+\alpha
\!\sum_{\mathbf{k}}k\left[ e^{i\varphi _{k}}\hat{a}_{{\mathbf{k}},\uparrow
}^{\dag }\hat{a}_{{\mathbf{k}},\downarrow }+h.c.\right]   \nonumber \\
&&\!\!\!+\frac{U_{b}}{L^{2}}\sum_{{\mathbf{k}},{\mathbf{k}}^{\prime },{%
\mathbf{q}}}\hat{a}_{{\mathbf{k}},\uparrow }^{\dag }\hat{a}_{-{\mathbf{k}}+{%
\mathbf{q}},\downarrow }^{\dag }\hat{a}_{-{\mathbf{k}}^{\prime }+{\mathbf{q}}%
,\downarrow }\hat{a}_{{\mathbf{k}}^{\prime },\uparrow }.  \label{eq:hamil}
\end{eqnarray}%
where $a_{{\mathbf{k}},\sigma }$ and $a_{{\mathbf{k}},\sigma }^{\dag }$
denote the fermionic field operators with in-plane wave vector ${\mathbf{k}}%
\equiv \left( k_{x},k_{y}\right) $ and spin $\sigma =\uparrow ,\downarrow $.
The free particle dispersion relation is given by $\xi _{{\mathbf{k}},\sigma
}=\epsilon _{\mathbf{k}}-\mu _{\sigma }$, where $\epsilon _{\mathbf{k}%
}=\hbar ^{2}\mathbf{k}^{2}/\left( 2m\right) $ and $\mu _{\uparrow \downarrow
}=\mu \pm h$, with $\mu $ being the chemical potential and $h$ being the effective
Zeeman field (equivalent to a population
imbalance between spin-up and spin-down components). The
second term of $H$ describes the Rashba-type SO coupling, whose strength is
denoted by the coefficient $\alpha $, with $k$ and $\varphi _{k}$ being the
magnitude and azimuthal angle of the the in-plane wave vector ${\mathbf{k}}$%
. A combination of the Zeeman field and the Rashba-type SO coupling can be
realized through control of a few laser beams \cite{4}. The $U_{b}$ term in
the Hamiltonian describes the effective 2D interaction, where $L$ is the
quantization length and $U_{b}$ is the bare coupling rate. The bare coupling
$U_{b}$ is connected with the physical coupling $U_{p}$ ($U_{p}$ is
determined by the 3D scattering length \cite{10}) through the 2D
renormalization relation~\cite{10}
\begin{equation}
U_{b}^{-1}=U_{p}^{-1}-L^{-2}\sum_{\mathbf{k}}\left( 2\epsilon _{\mathbf{k}%
}+\hbar \omega _{z}\right) ^{-1}.
\end{equation}

We determine the self-consistent phase diagram of the Hamiltonian $H$ within
the mean-field framework, which is a reasonable approximation at zero
temperature. Under this framework, we introduce the pairing order parameter $%
\Delta =U_{b}L^{-2}\sum_{\mathbf{k}}\left\langle \hat{a}_{-{\mathbf{k}}%
,\downarrow }\hat{a}_{{\mathbf{k}},\uparrow }\right\rangle $ to decompose
the last term of $H$ into $\left( \Delta \sum_{{\mathbf{k}}}\hat{a}_{{%
\mathbf{k}},\uparrow }^{\dag }\hat{a}_{-{\mathbf{k}},\downarrow }^{\dag
}+H.c.\right) -\left\vert \Delta \right\vert ^{2}L^{2}/U_{b}$ through the
Wick theorem. The quadratic mean-field Hamiltonian can then be diagonalized
into
\begin{equation}
H=\sum_{\mathbf{k}}\left( E_{{\mathbf{k}},+}\hat{\alpha}_{{\mathbf{k}}%
,+}^{\dag }\hat{\alpha}_{{\mathbf{k}},+}+E_{{\mathbf{k}},-}\hat{\alpha}_{{%
\mathbf{k}},-}^{\dag }\hat{\alpha}_{{\mathbf{k}},-}\right) +L^{2}\Omega _{0},
\label{eq:diag_hamil}
\end{equation}%
where $\hat{\alpha}_{{\mathbf{k}},\pm }$ denote the quasi-particle modes
with the excitation energies%
\begin{eqnarray}
E_{{\mathbf{k}},\pm }^{2}\!\!=\!\xi _{\mathbf{k}}^{2}\!+\!\alpha
^{2}k^{2}\!+\!h^{2}\!+\!\left\vert \Delta \right\vert ^{2}\!\!\pm\!\! \sqrt{\!\left( h^{2}\!+\!\alpha ^{2}k^{2}\right)\! \xi _{\mathbf{k}%
}^{2}\!+\!h^{2}\left\vert \Delta \right\vert ^{2}},
\end{eqnarray}%
and $\xi _{\mathbf{k}}=\epsilon _{\mathbf{k}}-\mu $. The last term in Eq.~(%
\ref{eq:diag_hamil}) denotes the zero-temperature thermodynamic potential,
with $\Omega _{0}$ given by
\begin{equation}
\Omega _{0}=L^{-2}\sum_{\mathbf{k}}\left[ \xi _{\mathbf{k}}-\left( E_{{%
\mathbf{k}},+}+E_{{\mathbf{k}},-}\right) /2\right] -\left\vert \Delta
\right\vert ^{2}/U_{b}.  \label{eq:Omega0}
\end{equation}%
At zero temperature, the quasi-particles $\hat{\alpha}_{{\mathbf{k}},\pm }$
are in vacuum states and we should minimize $\Omega _{0}$ with respect to
the order parameter $\Delta $ to determine the ground state of the system.

\begin{figure}[tbp]
\includegraphics[width=2.8in]{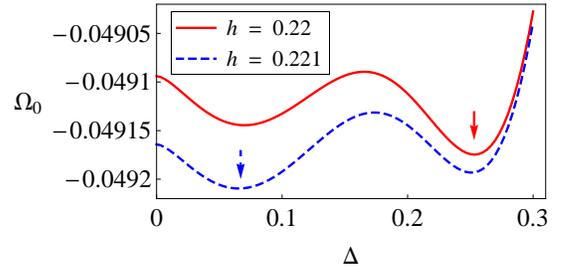}
\caption{The thermo-potential $\Omega _{0}$ as functions of the superfluid
order parameter $\Delta $ under different effective fields $h$. Other
parameters include the chemical potential $\protect\mu =0.5$, the spin-orbital
coupling strength $\protect\alpha =0.1$, and the interaction parameter $%
U_{p}=-5$ (corresponding to a 3D s-wave scattering length $a_s\approx -2 a_z$).
Normalized with the energy unit $\hbar \protect\omega _{z}$ and the length
unit $a_z$, all the parameters are dimensionless in this and the following figures. The arrows indicate the
global minimum of the corresponding curves.}
\label{fig:Omega0}
\end{figure}

Usually one calculates the phase diagram of the system by solving the gap
equation $\partial \Omega _{0}/\partial \Delta =0$ and the number equations%
\begin{equation}
n=n_{\uparrow }+n_{\downarrow },\quad n_{\sigma }=-\frac{\partial \Omega _{0}%
}{\partial \mu _{\sigma }}.
\end{equation}%
However, this is not a reliable approach here since the gap equation has
multiple solutions corresponding to unstable and metastable states. A
typical thermo-potential $\Omega _{0}$ is shown in Fig. \ref{fig:Omega0}. As
one can see, it has multiple minima and maxima satisfying $\partial \Omega
_{0}/\partial \Delta =0$. To find the ground state of the system, in this
paper we directly calculate $\Omega _{0}$ for various $\Delta $ to figure
out its global minimum. In our calculation, we take $\hbar \omega _{z}$ as
the energy unit and $a_{z}\equiv \sqrt{\hbar /\left( m\omega _{z}\right) }$
as the length unit. The units for $U_{p}$ and $\alpha $ are given
respectively by $a_{z}^{2}\hbar \omega _{z}$ and $a_{z}\hbar \omega _{z}$.
With these units, all the parameters become dimensionless.
\begin{figure}[tbp]
\includegraphics[width=2.5in]{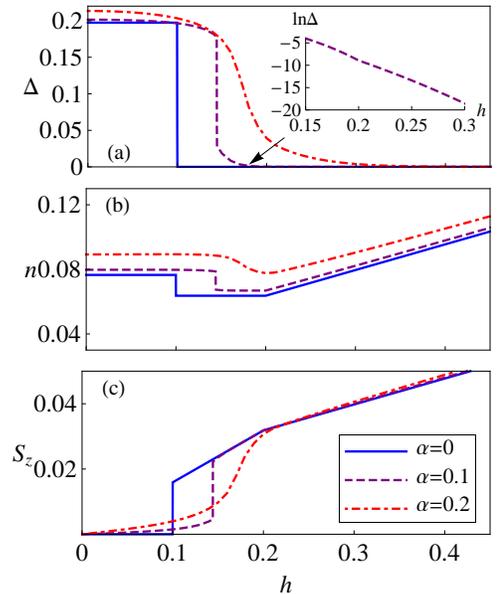}
\caption{The superfluid order parameter $\Delta$ (a), the number density $n$ (b), and the
magnetization $S_z\equiv(n_\uparrow-n_\downarrow)/2$ (c) as functions of the effective Zeeman field $h$ under
different spin-orbital coupling strengthes. The chemical potential and the
interaction parameter are fixed at $\protect\mu=0.2$ and ${U}_{p}=-5$. The
inset of Fig. (a) shows the exponential decay of the superfluid order
parameter with increase of the Zeeman field.}
\label{fig:Delta_h_1d}
\end{figure}
\begin{figure}[tbp]
\includegraphics[width=2.7in]{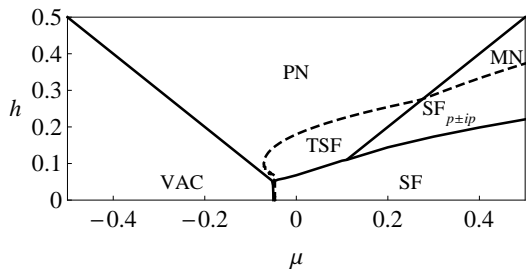}
\caption{The phase diagram of a polarized 2D Fermi gas under SO
coupling, with the parameters $U_p=-5$ and $\protect\alpha=0.1$. The SF,
TSF, SF$_{p \pm ip}$, VAC, represent, respecivtely, the conventional
superfluid, the topological superfluid, the spin-dependent mixture of $p_x
\pm ip_y$ superlfuid, and the vacuum phases. When the order parameter $%
\Delta $ gets extremely small, it is very likely that the superfluid will be
destroyed by thermal or quantum fluctuation, leading to normal mixture (MN)
or polarized normal (PN) phase. The dashed curve represents a crossover line
by setting $\Delta=10^{-6}$ (the choice of $10^{-6}$ is arbitrary,
however, the line has no significant change by setting a different small cutoff value for $\Delta$
due to its exponential decay). }
\label{fig:PhaseBoundary}
\end{figure}
\begin{figure}[tbp]
\includegraphics[width=3in]{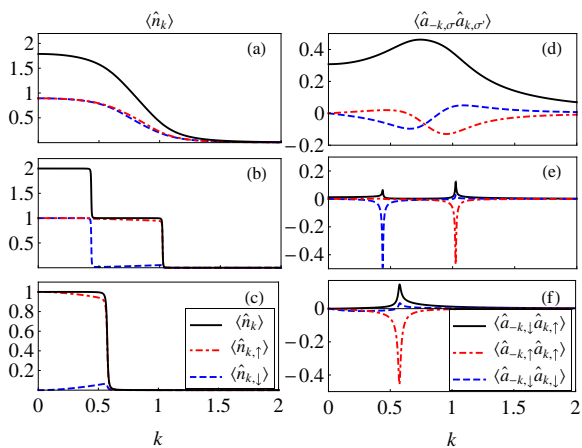}
\caption{The occupation number $n_k$ (a-c) and pairing functions $%
\left\langle\hat a_{-{\mathbf{k}},\protect\sigma}\hat a_{{\mathbf{k}},%
\protect\sigma^{\prime }}\right\rangle$ (d-f) as functions of $%
k_x$ for different phases: (a,d) the conventional
superfluid phase with parameters $\protect\mu=0.3$, $h=0.1$; (b,e) the
spin-dependent mixture of $p_x \pm ip_y$ superfluid phase with $\protect\mu%
=0.3$, $h=0.2$; (c,f) the topological superfluid phase with $\protect\mu%
=0.05 $, $h=0.1$.}
\label{fig:denk_k_1d}
\end{figure}

In Fig.~\ref{fig:Delta_h_1d}, we plot the superfluid order parameter $\Delta
$, the number density $n$, and the magnetization $S_{z}\equiv \left(
n_{\uparrow }-n_{\downarrow }\right) /2$, as functions of the effective
Zeeman field $h$ under various values of the SO coupling rate $\alpha $. At $%
\alpha =0$, it is clear there are two phase transitions induced by
increasing the field $h$. Below a critical field $h_{c}\sim 0.1$, we have a
superfluid phase with zero magnetization. This corresponds to the
conventional BCS\ state. Above $h_{c}$, the superfluid order parameter $%
\Delta $ is suppressed to zero, and we have a finite magnetization $S_{z}$
with $n_{\uparrow }>n_{\downarrow }>0$. This corresponds to the normal
mixture phase. Further increasing the field above $h_{c}\sim 0.2$, the
minority component $n_{\downarrow }$ is reduce to zero and we have a
polarized normal phase.

With a finite $\alpha $, the picture changes qualitatively. At a small $%
\alpha $, such as $\alpha =0.1$ in Fig. 2, there is still a big drop of the
the superfluid order parameter $\Delta $ when the field $h$ is above a
critical value $h_{c}$, but $\Delta $ does not drop to zero. The critical
field $h_{c}$ increase with $\alpha $, which indicates that the SO coupling
enhances superfluidity. The transition is still of the first order for small
$\alpha $ similar to the $\alpha =0$ case. What happens is that the
thermo-potential $\Omega _{0}$ as a function of $\Delta $ has two non-zero
minima with $\Delta _{2}>\Delta _{1}>0$, as shown in Fig. 1. As $h$
increases, $\Delta _{1}$, replacing $\Delta _{2}$, becomes the global
minimum, and the ground state undergoes a first-order transition between
different types of superfluid phases when the order parameter jumps from $%
\Delta _{2}$ to $\Delta _{1}$. Above the critical field $h_{c}$, the order
parameter $\Delta $ eventually decreases exponentially with increase of the
field $h$ as shown by the insert of Fig. 2(a), however, it does not reach
exact zero within the mean-field framework. Due to the exponential decrease
of $\Delta $, the order parameter becomes very small for large $h$ and it
will be destroyed by thermal fluctuation even at very low temperature or
likely by quantum fluctuation beyond the mean-field framework. At larger $%
\alpha $, the sudden drop of the order parameter $\Delta $ disappears, as
indicated by the $\alpha =0.2$ case in Fig. 2(a). If we look at the
derivative of $\Delta $ as a function of $h$ at $\alpha =0.2$, it shows a
kink at the critical $h_{c}$, suggesting that the phase transition changes
from the first order to the second order for large $\alpha $.

With nonzero SO coupling, the superfluid phase above the critical $h_{c}$
has novel features and is the focus of our interest in the following
discussion. We concentrate on the small $\alpha $ case, as it is easier to
realize this case in experiments. Below the critical $h_{c}$, the superfluid
phase is similar to the conventional BCS\ state, where the singlet-pairing
dominates. Above the critical $h_{c}$, there are two kinds of unconventional
superfluid phases. From the order parameter $\Delta $, it is hard to
distinguish these phases as $\Delta $ changes continuously across the phase
boundary. However, from the number density and the magnetization shown in
Fig. 2(b) and 2(c), one can clearly see a kink at the boundary, indicating a
second order phase transition. The boundary is determined by the condition $%
h^{2}=\mu ^{2}+\Delta ^{2}$. In different superfluid phases, the excitation
spectra $E_{{\mathbf{k}},\pm }$ of the quasiparticles are always gapped,
except at the critical point with $h=\sqrt{\mu ^{2}+\Delta ^{2}}$, where the
excitation spectrum $E_{{\mathbf{k}},-}$ becomes gapless. When $h>\sqrt{\mu
^{2}+\Delta ^{2}}$, the phase is identified with the topological superfluid
state which supports Majorana fermion excitations with exotic non-abelian
fractional statistics \cite{7,8}. When $h_{c}<h<\sqrt{\mu ^{2}+\Delta ^{2}}$%
, we have a new kind of unconventional superfluid phase whose nature will be
studied below.

The overall phase diagram is shown in Fig. 3 at $\alpha =0.1$. To understand
the nature of different superfluid phases, we show in Fig. 4 the
corresponding occupation number $\left\langle n_{{\mathbf{k}},\sigma
}\right\rangle $ in the momentum space and the pairing functions $%
\left\langle \hat{a}_{-{\mathbf{k}},\sigma }\hat{a}_{{\mathbf{k}},\sigma
^{\prime }}\right\rangle $. In the conventional superfluid phase denoted by
SF there, the pairing is dominantly in the s-wave channel between the spin $%
\uparrow $ and $\downarrow $ components with small hybridization from the
triplet pairing $\left\langle \hat{a}_{-{\mathbf{k}},\uparrow }\hat{a}_{{%
\mathbf{k}},\uparrow }\right\rangle $ and $\left\langle \hat{a}_{-{\mathbf{k}%
},\downarrow }\hat{a}_{{\mathbf{k}},\downarrow }\right\rangle $ induced by
the SO coupling. The momentum distribution $\left\langle n_{{\mathbf{k}}%
,\sigma }\right\rangle $ is rounded off by the strong pairing interaction
and $\left\langle n_{{\mathbf{k}},\uparrow }\right\rangle $ and $%
\left\langle n_{{\mathbf{k}},\downarrow }\right\rangle $ are almost
identical as the magnetization is small. In the topological superfluid phase
denoted by TSF there, the majority spin component $\left\langle n_{{\mathbf{k%
}},\uparrow }\right\rangle $ dominates. The pairing is mainly in the triplet
channel $\left\langle \hat{a}_{-{\mathbf{k}},\uparrow }\hat{a}_{{\mathbf{k}}%
,\uparrow }\right\rangle $ with a phase factor of $e^{i\varphi _{k}}$,\
where $\varphi _{k}$ denotes the azimuthal angle of ${\mathbf{k}}$. This
corresponds to the $p_{x}+ip_{y}$ superfluid phase, and it is well known
that this phase has Majorana Fermion excitations with non-abelian statistics
\cite{7,8}. For the superfluid phase with $h_{c}<h<\sqrt{\mu ^{2}+\Delta ^{2}%
},$ we have significant population in both spin components. The pairing is
dominantly in the triplet channel $\left\langle \hat{a}_{-{\mathbf{k}}%
,\uparrow }\hat{a}_{{\mathbf{k}},\uparrow }\right\rangle $ and $\left\langle
\hat{a}_{-{\mathbf{k}},\downarrow }\hat{a}_{{\mathbf{k}},\downarrow
}\right\rangle $. However, the symmetry is different for the $\uparrow
,\downarrow $ components: the pairing phase for $\left\langle \hat{a}_{-{%
\mathbf{k}},\uparrow }\hat{a}_{{\mathbf{k}},\uparrow }\right\rangle $ ($%
\left\langle \hat{a}_{-{\mathbf{k}},\downarrow }\hat{a}_{{\mathbf{k}}%
,\downarrow }\right\rangle $) is given by $e^{i\varphi _{k}}$ ($e^{-i\varphi
_{k}}$), suggesting that the spin-up and down components are in different
p-wave superfluid states, with $p_{x}+ip_{y}$ ($p_{x}-ip_{y}$) symmetries,
respectively. The system is in a mixture of these two spin-dependent $p\pm ip
$ states, and the phase is denoted by SF$_{p\pm ip}$ there. In a vortex of
this superfluid phase, both components may support Majorana fermions, but a
combination of them gives topologically trivial excitations. As a mixture of
two topological superfluids, we expect the SF$_{p\pm ip}$ phase has
topologically non-trivial properties, which could be activated through
spin-dependent manipulation and detection of the atomic cloud \cite{4}. This
deserves further study in future.

\begin{figure}[tbp]
\centering
\includegraphics[width=\linewidth]{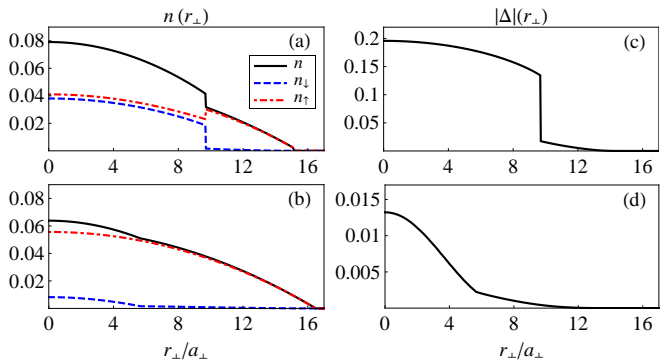}
\caption{The density profile (left column) and the superfluid order
parameter (right column) as functions of the radius $r_\perp$ of the atomic cloud
under different population imbalance: (a,c) $h=0.1$ and the chemical
potential at the trap center $\protect\mu_0\simeq0.20$, corresponding to the
total atom number $N=10^4$ and the polarization $P\equiv(N_\uparrow -
N_\downarrow)/N \simeq0.31$. (b,d) $\protect\mu_0=0.19$ and $h=0.15$,
corresponding to $N=10^4$ and $P=0.93$. The other parameters are $U_{p}=-5$,
$\protect\alpha=0.1$, and the aspect ratio $\protect\lambda=400$.}
\label{fig:lda}
\end{figure}

To have experimental signature of the transition between different
superfluid phases, we look at the density profile of the atomic gas in a
weak global harmonic trap, which can be measured directly in experiments
\cite{9}. Assume the global trap in the $x$-$y$ plane is given by $V_{\perp
}\left( x,y\right) =m\omega _{\perp }^{2}\left( x^{2}+y^{2}\right)
/2=m\omega _{\perp }^{2}r_{\perp }^{2}/2$ with the aspect ratio $\lambda
\equiv \omega _{z}/\omega _{\perp }\gg 1$. The density profile can be
calculated through the standard local density approximation which replaces $%
\mu $ in the homogeneous case by the position dependent $\mu \left( r_{\perp
}\right) =\mu _{0}-V_{\perp }\left( x,y\right) $, where $\mu _{0}$ denotes
the chemical potential at the trap center. The typical density profiles and
the corresponding superfluid order parameters are shown in Fig. 5 under two
scenarios. In Fig. 5(a) and 5(b), we see a first-order transition from the
conventional SF phase to the topological TSF phase, signified by a jump in
the density profile for both $\left\langle n_{{\mathbf{k}},\uparrow
}\right\rangle $ and $\left\langle n_{{\mathbf{k}},\downarrow }\right\rangle
$. In Fig. 5(c) and 5(d), we see a second-order phase transition from the SF$%
_{p\pm ip}$ phase to the TSF phase, indicated by a kink in the density
profile of $\left\langle n_{{\mathbf{k}},\downarrow }\right\rangle $.

In summary, we have self-consistently calculated the phase diagram of the 2D
polarized Fermi gas under SO coupling, and found two kinds of unconventional
superfluid phases with either topological non-abelian excitations or
spin-dependent mixing of $p\pm ip$ superfluid states. The transition order
between different superfluid phases is specified and the in-situ measurement
of the density profile of the atomic gas provides a convenient method to
reveal the phase transition and its order.

This work was supported by the NBRPC (973 Program) 2011CBA00300
(2011CBA00302), the DARPA OLE program, the IARPA MUSIQC program, the ARO and
the AFOSR MURI program.

\end{document}